\documentclass[a4paper, 11pt, reqno]{amsart}

\usepackage{graphicx}
\usepackage{appendix}
\usepackage{amsmath,amssymb,bm,fancybox}
\usepackage{xcolor} % arXiv用ではこのコメントアウトを外して、\usepackage[dvipdfmx]{color}をコメントアウトする
\usepackage{cite}
\usepackage{mathtools}
\usepackage{bm}
\usepackage{tikz}
\usetikzlibrary{arrows.meta}
\usepackage{algorithm}
\usepackage{algorithmic}

\usepackage{booktabs}
\usepackage{multirow}
\usepackage{longtable}
\usepackage{tabularx}
\usepackage{enumitem}

\newcommand{\bin}{\{0,1\}}
\newcommand{\ra}{\rightarrow}

\newcommand{\ol}[1]{\overline{#1}}

\newcommand{\B}{\clubsuit}
\newcommand{\R}{\heartsuit}
\newlength{\ralen}
\newcommand{\blk}{\raisebox{\ralen}{\framebox(11,13){$\B$}}}
\newcommand{\red}{\raisebox{\ralen}{\framebox(11,13){$\R$}}}

\newcommand{\back}{\raisebox{\ralen}{\framebox(11,13){\bf\large ?}}}

\newcommand{\bval}{b}%ビット値を表す変数

\begin{document}
\title{Impossibility Results of Card-Based Protocols via Mathematical Optimization}
\author{Shunnosuke Ikeda}
\address{University of Tsukuba, Japan.}
\email{ikeda@cs.tsukuba.ac.jp}
\author{Kazumasa Shinagawa}
\address{University of Tsukuba, Japan; Kyushu University, Japan; National Institute of Advanced Industrial Science and Technology (AIST), Japan.}
\email{shinagawa@cs.tsukuba.ac.jp}
\keywords{Card-based cryptography, combinatorics on words, mathematical optimization}
% \thanks{K. Shinagawa is supported by Japan Society for the Promotion of Science KAKENHI 21K17702 and 23H00479, JST K Program Grant Number JPMJKP24U2, and JST CREST Grant Number MJCR22M1.}

%
\maketitle              % typeset the header of the contribution
\begin{abstract}
This paper introduces mathematical optimization as a new method for proving impossibility results in the field of card-based cryptography. While previous impossibility proofs were often limited to cases involving a small number of cards, this new approach establishes results that hold for a large number of cards. The research focuses on single-cut full-open (SCFO) protocols, which consist of performing one random cut and then revealing all cards. The main contribution is that for any three-variable Boolean function, no new SCFO protocols exist beyond those already known, under the condition that all additional cards have the same color. The significance of this work is that it provides a new framework for proving impossibility results and delivers a proof that is valid for any number of cards, as long as all additional cards have the same color.
\end{abstract}
%\keywords{Card-based cryptography \and Combinatorics on words \and Mathematical optimization.}
%
%
%
\section{Introduction}\label{s:introduction}

It is known that cryptographic techniques such as secure multiparty computation and zero-knowledge proof can be physically represented using cards, and such a research area is called card-based cryptography \cite{BoerEC1989,KilianC1994,MizukiIJIS2014}.
Unlike ordinary cryptographic techniques, it is performed physically and visually, which makes them intuitive and provides a high degree of confidence in their security.
Due to this feature, its application to security education has been considered, and it has been used in lectures for cryptography \cite{Cheung2013,MarcedoneEPRINT2015,MizukiISEC2016,ShinagawaSCIS2022}.

An important research problem in this field is to determine which functions can be securely computed under specific conditions.
To achieve this, it is necessary to provide proofs that the function cannot be securely computed under specific conditions.
A representative method for proving impossibility results is the technique by Koch--Walzer--H\"{a}tel \cite{KochAC2015,KastnerAC2017,KochAC2019,KochMC2022}.
In this technique, a protocol is represented by a directed graph, and it is shown that no transition from a starting vertex to a terminal vertex is possible.
Using this method, impossibility proofs for AND protocols with five or fewer cards \cite{KochAC2015,KastnerAC2017,KochMC2022} and for COPY protocols of $k$ copies with $2k+1$ or fewer cards \cite{KastnerAC2017} have been obtained.
However, both of these results are in a setting where at most one additional card is used besides those required for the input and output.
In general, when the number of cards is large, a combinatorial explosion occurs. Therefore, to obtain further impossibility results, the development of new proof techniques seems to be necessary.

\subsection{Single-Cut Full-Open Protocols}\label{ss:SCFO_introduction}

% \begin{table}[t]
%  \centering
%  \caption{Summary of Existing SCFO Protocols for Boolean Functions}
%  \label{tab:SCFO}
%  \begin{tabular}{|c|c|c|c|c|} \hline
%  Authors & Function & \# Cards & Standard & $(k_0, k_1)$  \\ \hline
%  \multicolumn{5}{l}{$\circ$\, $2$-variable function} \\ \hline
%  Shinagawa et al.\cite{ShinagawaICISC2018} & $x_1 \oplus x_2$ & $4$ & $\checkmark$ & $(0,0)$  \\ \hline
%  den Boer~\cite{BoerEC1989} & $x_1 \wedge x_2$ & $5$ & $\checkmark$ & $(0,1)$  \\ \hline
%  \multicolumn{5}{l}{$\circ$\, $3$-variable function} \\ \hline
%  Heather et al.~\cite{HeatherFAOC2014} & $(x_1 = x_2 = x_3)?$ & $6$ & $\checkmark$ & $(0,0)$  \\ \hline
%  Shinagawa et al.~\cite{ShinagawaARXIV2025} & $(x_1 \wedge \ol{x_2}) \vee (x_2 \wedge \ol{x_3})$ & $8$ & $\checkmark$ & $(1,1)$  \\ \hline
%  Shinagawa et al.~\cite{ShinagawaARXIV2025} & $x_1 \oplus x_2 \oplus x_3$ & $8$ &  & --  \\ \hline
%  \multicolumn{5}{l}{$\circ$\, $4$-variable function} \\ \hline
%  Shinagawa et al.~\cite{ShinagawaARXIV2025} & $\ol{x_1}x_2\ol{x_4} \vee \ol{x_2}x_3\ol{x_4} \vee x_1\ol{x_2}x_4 \vee x_2\ol{x_3}x_4$ & $8$ & $\checkmark$ & $(0,0)$  \\ \hline
%  \end{tabular}
% \end{table}

\begin{table}[t]
\centering
\caption{Summary of Existing SCFO Protocols for Boolean Functions}
\label{tab:SCFO}
\begin{tabularx}{\textwidth}{llXccc}
\toprule
Category & Authors & Function & \#Cards & Standard & $(k_{0}, k_{1})$ \\
\midrule
\multirow{2}{*}{\textbf{2-variable}}
 & Shinagawa--Mizuki \cite{ShinagawaICISC2018} & $x_{1} \oplus x_{2}$ & 4 & $\checkmark$ & $(0, 0)$ \\
 & den Boer \cite{BoerEC1989} & $x_{1} \land x_{2}$ & 5 & $\checkmark$ & $(0, 1)$ \\
\midrule
\multirow{3}{*}{\textbf{3-variable}}
 & Heather et al. \cite{HeatherFAOC2014} & $(x_{1} = x_{2} = x_{3})?$ & 6 & $\checkmark$ & $(0, 0)$ \\
 & Shinagawa--Nuida \cite{ShinagawaARXIV2025} & $(x_1\wedge x_2) \vee (\ol{x_1} \wedge x_3)$ & 8 & $\checkmark$ & (1, 1) \\
 & Shinagawa--Nuida \cite{ShinagawaARXIV2025} & $x_{1}\oplus x_{2}\oplus x_{3}$ & 8 & -- & -- \\
\midrule
\multirow{2}{*}{\textbf{4-variable}}
 & \multirow{2}{*}{Shinagawa--Nuida \cite{ShinagawaARXIV2025}}
 & $\ol{x_1}x_2\ol{x_4} \vee \ol{x_2}x_3\ol{x_4} \vee $ & \multirow{2}{*}{8} & \multirow{2}{*}{$\checkmark$} & \multirow{2}{*}{$(0, 0)$} \\
 & & $x_1\ol{x_2}x_4 \vee x_2\ol{x_3}x_4$ & & & \\
%$\bar{x}_{1}x_{2}x_{3}\bar{x}_{4} \lor \bar{x}_{2}x_{3}x_{4} 
%   \lor x_{1}\bar{x}_{2}x_{4} \lor x_{2}x_{3}x_{4}$ 
\bottomrule
\end{tabularx}
\end{table}

A \emph{random cut} is a shuffling operation that cyclically and randomly shifts a sequence of cards. 
For a sequence of $k$ face-down cards $(c_0, c_1, \ldots, c_{k-1})$, a random cut is a probabilistic operation that chooses a uniformly random number $r \in \{0,1,\ldots,k-1\}$ and outputs a new sequence $(c_r, c_{r+1}, \ldots, c_{r+k-1})$, where the subscript is taken modulo $k$. 
Here, $r$ is hidden from all players. 
The following is an example for applying a random cut to a sequence of four cards:
\[
\overset{1}{\back}\,\overset{2}{\back}\,\overset{3}{\back}\,\overset{4}{\back}\,
\xrightarrow{~\mbox{random cut}~}\,
\begin{cases}
\overset{1}{\back}\,\overset{2}{\back}\,\overset{3}{\back}\,\overset{4}{\back}&\mbox{with probability $1/4$,}\vspace{3pt}\\
\overset{2}{\back}\,\overset{3}{\back}\,\overset{4}{\back}\,\overset{1}{\back}&\mbox{with probability $1/4$,}\vspace{3pt}\\
\overset{3}{\back}\,\overset{4}{\back}\,\overset{1}{\back}\,\overset{2}{\back}&\mbox{with probability $1/4$,}\vspace{3pt}\\
\overset{4}{\back}\,\overset{1}{\back}\,\overset{2}{\back}\,\overset{3}{\back}&\mbox{with probability $1/4$.}\vspace{3pt}
\end{cases}
\]

A \emph{single-cut full-open (SCFO) protocol} is a protocol of the following form:
\begin{enumerate}
\item Place a sequence of cards encoding an input. 
\item Turn over all face-up cards face-down, and then rearrange the sequence. 
\item Apply a random cut to the sequence of face-down cards.
\item Open all cards and determine an output value from the revealed symbols. 
\end{enumerate}
An SCFO protocol is \emph{standard} if the sequence in Step 1 is of the following form:
\[
\underset{x_1}{\back}\,\underset{\ol{x_1}}{\back}\,
\underset{x_2}{\back}\,\underset{\ol{x_2}}{\back}\,\cdots\,
\underset{x_n}{\back}\,\underset{\ol{x_n}}{\back}\,
%\underbrace{\back\,\back}_{x_1}\,\underbrace{\back\,\back}_{x_2}\,\cdots\,\underbrace{\back\,\back}_{x_n}\,
\underbrace{\blk\,\blk\,\cdots\,\blk}_{k_0\text{ cards}}\,\underbrace{\red\,\red\,\cdots\,\red}_{k_1\text{ cards}}\,,
\]
where a card with $x_i$ (resp. $\ol{x_i}$) represents a face-down card whose front side is $\blk$ if $x_i = 0$ (resp. $\ol{x_i} = 0$) and $\red$ otherwise. 
%the pair of cards encoding $x_i \in \bin$ is $\blk\,\red$ if $x_i = 0$ and $\red\,\blk$ if $x_i = 1$. 
A standard SCFO protocol using $k_0$ $\blk\,$s and $k_1$ $\red\,$s as additional cards is called a \emph{standard $(k_0,k_1)$-SCFO protocol}, and hereafter we simply call it a \emph{$(k_0,k_1)$-SCFO protocol}. 
When the value of $k_0$ is not specified, we refer to it as a $(*,k_1)$-SCFO protocol.
That is, a $(*,k_1)$-SCFO protocol uses $k_1$ $\red\,$s and any number of $\blk\,$s as additional cards.
Similarly, when the value of $k_1$ is not specified, we call it a $(k_0,*)$-SCFO protocol.

Table~\ref{tab:SCFO} shows a summary of existing SCFO protocols for Boolean functions. 
Except for the protocol for $x_1 \oplus x_2 \oplus x_3$, all protocols are standard. 
For $2$-variable Boolean functions, we can observe that any function can be computed by an SCFO protocol because AND and XOR protocols \cite{BoerEC1989,ShinagawaICISC2018} exist. 
However, for functions with $3$ or more variables, only a few can be computed by an SCFO protocol.
Whether there are other functions computable by SCFO protocols is an open problem.

\subsection{Our Contribution}\label{ss:contribution}

% \begin{table}[t]
%  \centering
%  \caption{Summary of Our Results for $(*,0)$-SCFO Protocols}
%  \label{tab:result}
%  \begin{tabular}{|c|c|c|} \hline
%  Function & ~~Existence of Protocol~~ & Protocol \\ \hline
%  \multicolumn{3}{l}{$\circ$\, $2$-variable function} \\ \hline
%  $x_1 \wedge x_2$ & $\checkmark$ & $(1,0)$-SCFO protocol~\cite{BoerEC1989} \\ \hline
%  $x_1 \oplus x_2$ & $\checkmark$ & ~~$(0,0)$-SCFO protocol~\cite{ShinagawaICISC2018}~~ \\ \hline
%  \multicolumn{3}{l}{$\circ$\, $3$-variable function} \\ \hline
%  $x_1 \wedge x_2 \wedge x_3$ & $\times$ & --- \\ \hline
%  $x_1 \oplus x_2 \oplus x_3$ & $\times$ & ---\\ \hline
%  $(x_1 = x_2 = x_3)?$ & $\checkmark$ & $(0,0)$-SCFO protocol~\cite{HeatherFAOC2014}\\ \hline
%  $(x_1 + x_2 + x_3 \geq 2)?$ & $\times$ & ---\\ \hline
%  $(x_1 + x_2 + x_3 = 1)?$ & $\times$ & ---\\ \hline
%  $x_1\oplus (x_2 \wedge x_3)$ & $\times$ & ---\\ \hline
%  ~~$(x_1 \wedge \ol{x_2}) \vee (x_2 \wedge \ol{x_3})$~~ & $\times$ & ---\\ \hline
%  $x_1\wedge (x_2 \oplus x_3)$ & $\times$ & ---\\ \hline
%  $(x_1\vee x_2) \wedge (x_1 \oplus x_3)$ & $\times$ & ---\\ \hline
%  $x_1\wedge (x_2 \vee x_3)$ & $\times$ & ---\\ \hline
%  \end{tabular}
%  %\\
%  %Note that the protocol for $x_1 \wedge x_2$ in \cite{BoerEC1989} is a $(0,1)$-SCFO protocol, \\but it implies a $(1,0)$-SCFO protocol by symmetry of $\clubsuit$ and $\heartsuit$.
% \end{table}

\begin{table}[t]
\centering
\caption{Summary of Our Results for $(*,0)$-SCFO Protocols}
\label{tab:result}
\begin{tabularx}{\textwidth}{llcc}
\toprule
Category & Function & Existence of Protocol & Protocol \\
\midrule
\multirow{2}{*}{\textbf{2-variable}}
 & $x_1 \wedge x_2$ & $\checkmark$ & $(1,0)$-SCFO protocol~\cite{BoerEC1989} \\
 & $x_1 \oplus x_2$ & $\checkmark$ & $(0,0)$-SCFO protocol~\cite{ShinagawaICISC2018} \\
\midrule
\multirow{9}{*}{\textbf{3-variable}}
 & $x_1 \wedge x_2 \wedge x_3$ & $\times$ & --- \\
 & $x_1 \oplus x_2 \oplus x_3$ & $\times$ & --- \\
 & $(x_1 = x_2 = x_3)?$ & $\checkmark$ & $(0,0)$-SCFO protocol~\cite{HeatherFAOC2014} \\
 & $(x_1 + x_2 + x_3 \geq 2)?$ & $\times$ & --- \\
 & $(x_1 + x_2 + x_3 = 1)?$ & $\times$ & --- \\
 & $x_1 \oplus (x_2 \wedge x_3)$ & $\times$ & --- \\
 & $(x_1 \wedge \overline{x_2}) \vee (x_2 \wedge \overline{x_3})$ & $\times$ & --- \\
 & $x_1 \wedge (x_2 \oplus x_3)$ & $\times$ & --- \\
 & $(x_1 \vee x_2) \wedge (x_1 \oplus x_3)$ & $\times$ & --- \\
 & $x_1 \wedge (x_2 \vee x_3)$ & $\times$ & --- \\
\bottomrule
\end{tabularx}
\end{table}

% \item $x_1 \wedge x_2 \wedge x_3$ (AND function)
% \item $x_1 \oplus x_2 \oplus x_3$ (XOR function)
% \item $(x_1 = x_2 = x_3)?$ (EQ function)
% \item $(x_1 + x_2 + x_3 \geq 2)?$ (MAJ function)
% \item $(x_1 + x_2 + x_3 = 1)?$
% \item ~$x_1\oplus (x_2 \wedge x_3)$
% \item ~$(x_1\wedge x_2) \vee (\ol{x_1} \wedge x_3)$
% \item ~$x_1\wedge (x_2 \oplus x_3)$
% \item ~$(x_1\vee x_2) \wedge (x_1 \oplus x_3)$
% \item ~$x_1\wedge (x_2 \vee x_3)$

In this paper, we propose a new technique for proving impossibility results in card-based cryptography, based on mathematical optimization.
This is the first work to apply mathematical optimization to card-based cryptography, and an important contribution of our work is demonstrating the effectiveness of mathematical optimization in this field.
Our concrete contributions are as follows:
\begin{itemize}
\item We formulated the search for a $(*,0)$-SCFO protocol as an integer optimization problem and developed an algorithm that takes a function as input and determines whether a protocol exists and, if so, outputs a protocol with the minimum number of additional cards.
\item We implemented this algorithm and conducted experiments for all $3$-variable Boolean functions (see Section \ref{ss:three_variable_function}). As a result, we obtained an impossibility proof for a $(*,0)$-SCFO protocol for all $3$-variable functions except for the $3$-variable equality function (see Table \ref{tab:result}). 
\end{itemize}

\subsection{Existing Methods for Proving Impossibility Results}\label{ss:relatedwork}

This section introduces existing methods for proving impossibility results in the field of card-based cryptography.
The technique proposed in this paper is a new proof method that does not belong to any of these classifications.

\begin{itemize}
\item \textbf{Information-theoretic proofs:} Mizuki--Shizuya~\cite{MizukiIJIS2014} proved that no perfect COPY protocol exists for single-card encoding. Intuitively, this means that in a protocol with single-card encoding, it is necessary to reveal the input card to satisfy correctness, but doing so leaks information, making correctness and security mutually exclusive.
\item \textbf{Proofs based on KWH diagrams:} Koch--Walzer--H\"{a}rtel~\cite{KochAC2015} proposed a diagram to visually represent protocols and, based on it, proved the impossibility of a (committed-format) 4-card AND protocol with finite-runtime. Although many results have been proven in follow-up studies~\cite{KastnerAC2017,KochAC2019}, this method is difficult to apply when the number of cards is large, and all previous results are for cases with zero or one additional cards.
\item \textbf{Proofs based on formal verification:} Koch--Schrempp--Kirsten~\cite{KochAC2019,KochNGCO2021} and its subsequent work~\cite{FujitaAPKC2025} proved impossibility results using formal verification. This method is also difficult to apply when the number of cards is large, and its impossibility results are for cases with four or fewer cards.
\item \textbf{Proofs based on the ABC conjecture:} Hashimoto et al.~\cite{HashimotoIEICE2018a} proved that $\Omega(n \log n)$ cards are required for the uniform random generation of a derangement of order $n$. The proof is based on the integer factorization of the number of derangements and uses the ABC conjecture.
\item \textbf{Proofs based on PSM protocols:} Shinagawa--Nuida~\cite{ShinagawaSTACS2025} proposed a general method for converting card-based protocols into PSM protocols. In particular, a single-shuffle full-open (SSFO) protocol with $n$ inputs and $k$ cards can be converted into a PSM protocol with communication complexity $nk$. This shows that a lower bound on the number of cards for an SSFO protocol can be derived from a lower bound on communication complexity of the PSM protocol.
\end{itemize}

\section{Preliminaries}

In Section \ref{ss:three_variable_function}, we introduce 3-variable Boolean functions.
In Section \ref{ss:cyclically_equalizable}, we define \emph{cyclically equal} \cite{ShinagawaFCT2025} from combinatorics on words.
Sections \ref{ss:card} to \ref{ss:SCFO} provide the basic definitions from card-based cryptography.

% Throughout this paper, we denote the set of positive integers from 1 to $n$ as $[n]:=\{1,2,\dots,n\}$.
% We denote the symmetric group of degree $n$ by $\mathfrak{S}_n$.
For a finite set $A$, we denote the set of finite-length sequences of $A$ by $A^*$. 
For example, $\{0\}^* = \{\epsilon, 0, 00, 000, 0000, \ldots\}$ and $\bin^* = \{\epsilon, 0, 1, 00, 01, 10, 11, \ldots\}$, where $\epsilon$ is an empty sequence. 

\subsection{NPN-Equivalence of 3-Variable Boolean Functions}\label{ss:three_variable_function}

For an $n$-variable Boolean function, the equivalence class obtained by a combination of the following three operations is called an NPN-equivalence class.
\begin{itemize}
\item Negation of some or all of the input variables
\item Permutation of the $n$ input variables
\item Negation of the output
\end{itemize}
When functions $f, g: \bin^n \ra \bin$ belong to the same NPN-equivalence class, if a protocol for $f$  exists, then a protocol for $g$ also exists.
Therefore, to perform secure computation for any $n$-variable Boolean function, it is sufficient to construct protocols for representatives of the NPN-equivalence classes. 

There are 14 representatives for the NPN-equivalence classes of 3-variable Boolean functions as follows:
\begin{enumerate}\renewcommand{\labelenumi}{\arabic{enumi}.}
\item $0$ (constant function)
\item $x_1$
\item $x_1 \wedge x_2$
\item $x_1 \oplus x_2$
\item $x_1 \wedge x_2 \wedge x_3$
\item $x_1 \oplus x_2 \oplus x_3$
\item $(x_1 = x_2 = x_3)?$
\item $(x_1 + x_2 + x_3 \geq 2)?$
\item $(x_1 + x_2 + x_3 = 1)?$
\item $x_1\oplus (x_2 \wedge x_3)$
\item $(x_1\wedge x_2) \vee (\ol{x_1} \wedge x_3)$
\item $x_1\wedge (x_2 \oplus x_3)$
\item $(x_1\vee x_2) \wedge (x_1 \oplus x_3)$
\item $x_1\wedge (x_2 \vee x_3)$
\end{enumerate}

\subsection{Cyclically Equal}\label{ss:cyclically_equalizable}

Two words $w,w' \in \bin^n$ are cyclically equal if there exist $u,v \in \bin^*$ such that $w = uv$ and $w' = vu$, where $uv$ and $vu$ represent the concatenation of words.
We write $w \sim w'$ if $w$ and $w'$ are cyclically equal, and $w \not\sim w'$ otherwise.
For example, for $w = 10110$ and $w' = 11010$, we can take $u = 10$ and $v = 110$, thus $w \sim w'$.
A set of words $\bm{w} \subseteq \bin^n$ is said to be cyclically equal if $w \sim w'$ holds for any $w,w' \in \bm{w}$.

For a set of words $\bm{w} \subseteq \bin^n$, an insertion is defined as an operation that transforms each word $w = w_1 w_2 \cdots w_n \in \bm{w}$ into a word $w' = w_1 u_1 w_2 u_2 \cdots w_n u_n$ for some common words $u_1,\dots,u_n \in \bin^*$, where each $u_i$ may be the empty word. 
We denote an insertion by $\bm{u} = (u_1, \ldots, u_n)$, and we write $w \xrightarrow{\bm{u}} w'$ to denote that the word $w$ is transformed into the word $w'$ by the simultaneous insertion $\bm{u}$.
In particular, a $0$-insertion is defined as an insertion such that all $u_1,\dots,u_n$ are repetitions of $0$, i.e., $u_i \in \{0\}^*$. 
Similarly, a $1$-insertion is defined as an insertion such that all $u_1,\dots,u_n$ are repetitions of $1$.

\subsection{Cards and Encoding}\label{ss:card}

We use a set of cards with two types of faces $\blk$ and $\red\,$, and identical backs $\back\,$.
In the Mizuki--Shizuya model~\cite{MizukiIJIS2014}, face-up cards are denoted by $\clubsuit/?$ or $\heartsuit/?$, and face-down cards are denoted by $?/\clubsuit$ or $?/\heartsuit$. For example, the following sequence
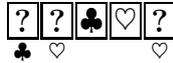
\[
\underset{\clubsuit}{\back}\,\underset{\heartsuit}{\back}\,\blk\,\red\,\underset{\heartsuit}{\back}
\]
is represented as $(?/\clubsuit,?/\heartsuit,\clubsuit/?,\heartsuit/?,?/\heartsuit)$.
In this paper, we represent a sequence of cards such as $\clubsuit\heartsuit\clubsuit\heartsuit\heartsuit$ since we only deal with face-up or face-down sequences, so the context determines whether a sequence is face-up or face-down without ambiguity.
We also denote the set of all sequences of $n$ cards by $\{\clubsuit,\heartsuit\}^n$.

In this paper, we let $\clubsuit = 0$ and $\heartsuit = 1$. That is, we identify the set of all sequences of $n$ cards $\{\clubsuit,\heartsuit\}^n$ with the set of all $n$-bit words $\bin^n$; for example, $\clubsuit\clubsuit\heartsuit\heartsuit\heartsuit = 00111$.

For a bit value $x \in \bin$, a pair of 
%two 
face-down cards $(x,\ol{x})$ is called a commitment to $x$.
That is, it is $(\clubsuit,\heartsuit)$ when $x = 0$, and $(\heartsuit,\clubsuit)$ when $x = 1$.
A commitment to $x$ is represented as follows:
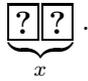
\[
\underbrace{\back\,\back}_{x}\,.
\]
In card-based cryptography, we typically consider situations where a sequence of commitments is given as input; thus, at least $2n$ cards are used for an $n$-bit input. It is known that perfect security cannot be achieved if the single-card encoding~\cite{NiemiTCS1998,ShinagawaICTAC2024} representing each bit with a single card is used~\cite{MizukiIJIS2014}. Therefore, single-card encoding is not typically used.

\subsection{Random Cut}\label{ss:randomcut}

A random cut is a shuffling operation that performs a cyclic shift on a sequence of cards a random number of times (see also Section \ref{ss:SCFO_introduction}).
% For a sequence of $n$ cards $w =w_1 w_2 \cdots w_n \in \{\clubsuit,\heartsuit\}^n$, we define the cyclic shift operation as $\sigma(w) = w_2 w_3 \cdots w_n w_1$.
% A random cut is an operation that applies $\sigma^r$ to a sequence of cards, where $0 \leq r \leq n-1$ is a uniform random number.
% Here, the random number $r$ is hidden from all players, so no one can guess it.
%In other words, 
When a random cut is applied to a sequence of $n$ cards $w =w_1 w_2 \cdots w_n \in \{\clubsuit,\heartsuit\}^n$, a word is chosen uniformly at random from the set of all words that are cyclically equal to $w$. 
Therefore, if $w_0 \sim w_1$ holds, the probability distribution of the resulting sequence of cards when a random cut is applied to $w_0$ is identical to the probability distribution of the resulting sequence when a random cut is applied to $w_1$.

For example, if a random cut is applied to a face-down sequence of five cards $\clubsuit\clubsuit\heartsuit\heartsuit\heartsuit$, the five possible sequences of cards $\clubsuit\clubsuit\heartsuit\heartsuit\heartsuit$, $\clubsuit\heartsuit\heartsuit\heartsuit\clubsuit$, $\heartsuit\heartsuit\heartsuit\clubsuit\clubsuit$, $\heartsuit\heartsuit\clubsuit\clubsuit\heartsuit$, and $\heartsuit\clubsuit\clubsuit\heartsuit\heartsuit$ will each occur with probability $1/5$, and no one can guess which sequence was actually chosen.

It is known that a random cut can be easily implemented by human hand. Furthermore, a method of implementing a random cut by placing cards on a spinning top and performing a rotation operation is also known.

\subsection{Single-Cut Full-Open Protocols}\label{ss:SCFO}

In this section, we introduce the notation used in SCFO protocols for subsequent sections. See Section \ref{ss:SCFO_introduction} for the definition of SCFO protocols.
Let $\mathfrak{S}_n$ denote the symmetric group of degree $n$.

A $(k_0,k_1)$-SCFO protocol proceeds as follows:
\begin{enumerate}\renewcommand{\labelenumi}{\arabic{enumi}.}
\item Arrange the input sequence as $x_1\ol{x_1}\cdots x_n\ol{x_n} \in \{\clubsuit,\heartsuit\}^{2n}$.
\item Permute the sequence with a permutation $\pi \in \mathfrak{S}_{2n}$.
\item Insert $k_0$ $\blk\,$s and $k_1$ $\red\,$s at arbitrary positions in the sequence.
\item Apply a random cut to the sequence.
\item Open all cards.
\end{enumerate}

Let $f: \bin^n \ra \bin$ be a function. 
For each $\bval \in \bin$, let $\bm{v}^{(\bval)}$ be the set of input sequences at Step 1 corresponding to the output value $\bval$ defined as:
\begin{equation}
    \bm{v}^{(\bval)} := \{\pi(x_1\ol{x_1}x_2\ol{x_2}\cdots x_n\ol{x_n}) \mid x_i \in \bin, f(x_1, \ldots, x_n) = b\}.\label{eq:def_v}
\end{equation}
% hereinafter $\pi(\cdot)$ denotes a permuting a sequence, i.e., $\pi(c_1c_2\cdots c_k) = c_{\pi^{-1}(1)}c_{\pi^{-1}(2)}\cdots c_{\pi^{-1}(k)}$ for $\pi \in \mathfrak{S}_k$. 
Let $\bm{w}^{(\bval)}$ be the set of sequences at Step 2 corresponding to $\bval$ defined as:
\begin{equation}
\bm{w}^{(\bval)} := \{\pi(v) \mid v \in \bm{v}^{(\bval)}\}.\label{eq:def_w}  
\end{equation}
Let $\bm{w}'^{(\bval)}$ be the set of sequences at Step 3 corresponding to $\bval$ defined as:
\begin{equation}
\bm{w}'^{(\bval)} := \{w' \mid w \in \bm{w}^{(\bval)}, w \xrightarrow{\bm{u}} w'\},\label{eq:def_w_prime}
\end{equation}
where $\bm{u} \in (\bin^*)^{2n}$ denotes the insertion at Step 3. 
The above protocol correctly and securely computes $f$ if it satisfies the following two conditions:
\begin{description}
\item[Correctness.] For any $w_0 \in \bm{w}'^{(0)}$ and $w_1 \in \bm{w}'^{(1)}$, we have $w_0 \not\sim w_1$.
\item[Security.] For each $\bval \in \bin$ and any $w_b, w'_b \in \bm{w}'^{(b)}$, we have $w_b \sim w'_b$. 
\end{description}

\section{Searching for $(*,0)$-SCFO Protocols}
In this section, we propose a method to search for $(*,0)$-SCFO protocols based on mathematical optimization.
To search for $(*,0)$-SCFO protocols, we need to find a permutation $\pi \in \mathfrak{S}_{2n}$ and a $0$-insertion $\bm{u} \in (\{0\}^*)^{2n}$ that yield $\bm{w}'^{(\bval)}$ satisfying the correctness and security conditions in Section~\ref{ss:SCFO}.

Our search method is based on an integer optimization formulation for finding a 0-insertion $\bm{u}$. 
The overall search procedure consists of the following two steps:
\begin{enumerate}[label=Step \arabic*., leftmargin=*, align=left]
\item For each permutation $\pi \in \mathfrak{S}_{2n}$, we compute $\bm{w}^{(\bval)}$ of Eq.~(\ref{eq:def_w}), and then solve an integer optimization problem to find a $0$-insertion $\bm{u}$ such that $\bm{w}'^{(\bval)}$ of Eq.~(\ref{eq:def_w_prime}) is cyclically equal, which corresponds to the security condition. The objective is to minimize the bit length of $\bm{u}$. This step returns a minimal-length insertion if a solution exists.
%\item The above step (Step 1) is executed for all permutations $\pi \in \mathfrak{S}_{2n}$ to identify a pair $(\pi, \bm{u})$ that satisfies the security requirements.
\vspace{2mm}
\item If we find a pair $(\pi, \bm{u})$ in the above step, it is then verified that $w_0 \in \bm{w}'^{(0)}$ and $w_1 \in \bm{w}'^{(1)}$ are not cyclically equal, which corresponds to the correctness. If the correctness is satisfied, the pair is output as a valid solution.
\end{enumerate}
The problem formulation used in Step 1 is detailed in Sections~\ref{ss:notation} and \ref{ss:formulation}.
Note that the optimization problem in Step 1 requires not only a permutation $\pi$ but also shift amounts as input, which are described later.
We address Step 2 in Section~\ref{ss:algorithm}.
Using this algorithm, we demonstrate the impossibility of a $(*,0)$-SCFO protocol for three-variable Boolean functions.
Furthermore, due to the symmetry between $\blk$ and $\red\,$, our result also implies the impossibility of a $(0,*)$-SCFO protocol for three-variable Boolean functions.

\subsection{Notation} \label{ss:notation}

Suppose that $\bm{w}^{(\bval)}$ of Eq.~(\ref{eq:def_w}) is given for each $\bval \in \bin$. 
In this subsection and the next subsection, we search for the insertion $\bm{u} = (u_1, u_2, \ldots, u_{2n}) \in (\{0\}^*)^{2n}$ that yields $\bm{w}'^{(\bval)}$ of Eq.~(\ref{eq:def_w_prime}) satisfying the correctness and security in Section~\ref{ss:SCFO}. 

We denote the set of positive integers from 1 to $n$ by $[n]:=\{1,2,\dots,n\}$.
Let $K^{(\bval)}$ denote the number of elements of $\bm{w}^{(\bval)}$. 
Since the number of elements of $\bm{w}^{(0)} \sqcup \bm{w}^{(1)}$, the disjoint union of $\bm{w}^{(0)}$ and $\bm{w}^{(1)}$, is equal to the number of all inputs, we have $K^{(0)} + K^{(1)} = 2^n$. 
Define $I := 2n$ and $J := n$. 
Here, $I$ represents the bit length of sequences, and $J$ represents the number of 1s in each sequence. 

Let $w^{(\bval)}_k$ be the $k$-th sequence of $\bm{w}^{(\bval)}$, i.e., $\bm{w}^{(\bval)} = \{w^{(\bval)}_1, \ldots, w^{(\bval)}_{K^{(\bval)}}\}$. 
Let $t^{(\bval)}_{j,k} \in [I]$ for $j \in [J]$ be the position of the $j$-th $1$ from the left. 
If we treat the 1s as \lq\lq separators\rq\rq{}, the sequence can be regarded as consisting of $J$ segments by considering the leftmost and the rightmost segments are connected. 
For example, an $8$-bit sequence $01100110$ consists of four segments as follows:
\[
\underbrace{\phantom{00}0\phantom{00}}_{\mbox{1st}}~1~\underbrace{\phantom{00000}}_{\mbox{2nd}}~1~\underbrace{\phantom{0}0\phantom{0}0\phantom{0}}_{\mbox{3rd}}~1~\underbrace{\phantom{00000}}_{\mbox{4th}}~1~\underbrace{\phantom{00}0\phantom{00}}_{\mbox{1st}}.
\]
Then the number of 0s in the $j$-th segment ($j \in [J]$) is given by
\begin{eqnarray*}
x^{(\bval)}_{j, k}
=
 \begin{cases}
 (t^{(\bval)}_{j,k} - 1) + (I - t^{(\bval)}_{J,k}) & ( j = 1 ), \\
t^{(\bval)}_{j,k} - t^{(\bval)}_{j-1,k} - 1 & ( 1 < j \le J ).
\end{cases}
\end{eqnarray*}

We now consider the insertion. 
For an $I$-bit sequence, there are $I$ possible insertion positions. 
For example, $01100110$ has 8 possible insertion positions as follows.
\begin{center}
\begin{tikzpicture}[baseline=(current bounding box.center)]
% 座標の設定
\node (v1) at (0.0,0) {0};
\node (v2) at (0.5,0) {1};
\node (v3) at (1.0,0) {1};
\node (v4) at (1.5,0) {0};
\node (v5) at (2.0,0) {0};
\node (v6) at (2.5,0) {1};
\node (v7) at (3.0,0) {1};
\node (v8) at (3.5,0) {0};
% pラベル
% \node (p0) at (-0.5, 0.8) {};
% \node (p1) at (0.5, 0.8) {};
% \node (p2) at (1.5, 0.8) {};
% \node (p3) at (2.5, 0.8) {};
% \node (p4) at (3.5, 0.8) {};
% 矢印
\draw[->] (0.25,0.5) -- (0.25,0.2);
\draw[->] (0.75,0.5) -- (0.75,0.2);
\draw[->] (1.25,0.5) -- (1.25,0.2);
\draw[->] (1.75,0.5) -- (1.75,0.2);
\draw[->] (2.25,0.5) -- (2.25,0.2);
\draw[->] (2.75,0.5) -- (2.75,0.2);
\draw[->] (3.25,0.5) -- (3.25,0.2);
\draw[->] (3.75,0.5) -- (3.75,0.2);
% 挿入ポジション
\node (l1) at (0.25,0.7) {1};
\node (l2) at (0.75,0.7) {2};
\node (l3) at (1.25,0.7) {3};
\node (l4) at (1.75,0.7) {4};
\node (l5) at (2.25,0.7) {5};
\node (l6) at (2.75,0.7) {6};
\node (l7) at (3.25,0.7) {7};
\node (l8) at (3.75,0.7) {8};
\end{tikzpicture}
%\vspace{2mm}
\end{center}
For a sequence $w^{(\bval)}_k \in \bin^I$, define a constant $a^{(\bval)}_{i,j,k} \in \bin$ to indicate whether an insertion position $i$ belongs to a segment $j$:
\[
a^{(\bval)}_{i,j,k}
=
\begin{cases}
1&\text{if position $i$ belongs to segment $j$},\\
0&\text{otherwise},
\end{cases}
\]
for all $i\in[I],\,j\in[J],\,k\in[K^{(\bval)}],\,\text{and} ~\bval\in\{0,1\}$.
Let $y_i \in \mathbb{Z}_{\ge 0}~(i \in [I])$ be the number of 0s inserted at position $i$. 
%, i.e., the insertion vector $\bm{u}$ can be written by $(0^{y_1}, 0^{y_2}, \ldots, 0^{y_{2n}})$
After the insertion, the number of 0s for the $j$-th segment in $w^{(\bval)}_k$ is given by:
\[
x'^{(\bval)}_{j,k}
\;=\;
x^{(\bval)}_{j,k}
\;+\;\sum_{i\in[I]}a^{(\bval)}_{i,j,k}\,y_i\,.
\]
Then, two sequences $w^{(\bval)}_k$ and $w^{(\bval)}_{k+1}$ will become cyclically equal by the insertion if the vector of integers $(x'^{(\bval)}_{1,k}, \ldots, x'^{(\bval)}_{J,k})$ is equal to a cyclic shift of the vector of integers $(x'^{(\bval)}_{1,k+1}, \ldots, x'^{(\bval)}_{J,k+1})$. 
In other words, if the shift amount is $s^{(\bval)}_{k+1} \in \{0, 1, \ldots, J-1\}$, the following equation holds for all $j \in [J]$:
\begin{align}   
{x'}_{j,k}^{(\bval)} = {x'}^{(\bval)}_{\bigl((j + s^{(\bval)}_{k+1} - 1) ~\mathrm{mod} ~J \bigr)+1,k+1}. \label{eq:cyclic_equal}
\end{align}
%where $\bm{x}'^{(\bval)}_{k} \coloneqq (x'^{(\bval)}_{j,k})_{j \in [n]} \in \mathbb{Z}^{n}_{\ge 0}$.

% Specifically, let $w^{(1)}_1 = 0110$ and $w^{(1)}_2 = 1001$.
% Without 0-insertions, their segment length vectors are $\bm{x}'^{(1)}_1 = (2, 0)$ and $\bm{x}'^{(1)}_2 = (0, 2)$.
% % based on the following calculations:
% % \[
% % {x'}^{(1)}_{1,1} = {x}^{(1)}_{1,1} = 2, \quad {x'}^{(1)}_{2,1} = {x}^{(1)}_{2,1} = 0,
% % \]
% % \[
% % {x'}^{(1)}_{1,2} = {x}^{(1)}_{1,2} = 0, \quad {x'}^{(1)}_{2,2} = {x}^{(1)}_{2,2} = 2\,,
% % \]
% For the shift amount $s^{(1)}_2 = 1$, the cyclic equality condition \eqref{eq:cyclic_equal} is satisfied:
% \[
% {x'}^{(1)}_{\bigl((1 + s^{(1)}_2 - 1) ~\mathrm{mod} ~2\bigr)+1, 2} = {x'}^{(1)}_{2, 2} = 2 = {x'}^{(1)}_{1,1},
% \]
% \[
% {x'}^{(1)}_{\bigl((2 + s^{(1)}_2 - 1) ~\mathrm{mod} ~2\bigr)+1, 2} = {x'}^{(1)}_{1, 2} = 0 = {x'}^{(1)}_{2,1}.
% \]
% Thus, $w^{(1)}_1$ and $w^{(1)}_2$ are cyclically equal in this case.

\subsection{Problem Formulation} \label{ss:formulation}

Given a set of input sequences $\bm{w} \coloneqq \bm{w}^{(0)} \sqcup \bm{w}^{(1)}$, the search for a $(*,0)$-SCFO protocol consists of finding the following two vectors:
\begin{itemize}
\item $0$-insertion vector: $\bm{y} \coloneqq (y_i)_{i \in [I]} \in \mathbb{Z}^{I}_{\ge 0}$.
\item Shift amount vector: $\bm{s} \coloneqq (s^{(\bval)}_{k})_{\bval \in \bin, k \in \{2,3,\ldots, K^{(\bval)}\}}\in \mathbb{Z}_{\ge 0}^{K^{(0)} + K^{(1)} - 2}$.
\end{itemize}
Here, the index $k$ of the shift amount vector starts from $2$ because the shift amount $s^{(\bval)}_{k}$ represents the shift amount of $w_k^{(\bval)}$ relative to $w_1^{(\bval)}$ for $2 \leq k \leq K^{(\bval)}$.
These vectors must be determined such that all sequences corresponding to the same output value become cyclically equal after the 0-insertion.

To achieve this, we formulate the search as an optimization problem.
Here, we consider the shift amount vector $\bm{s}$ to be a fixed parameter, and seek an insertion vector $\bm{y}$ that achieves cyclic equality with the minimum number of inserted 0s.
This leads to the following integer optimization problem, denoted $\mathrm{P}(\bm{s, w})$.\\\\
$\mathrm{P}(\bm{s, w}):$
\begin{align}
\underset{\bm{x}', \bm{y}~~}{\parbox{5em}{minimize}} \quad & \sum_{i \in [I]} y_i \label{eq:obj} \\
\parbox{5em}{subject to} \quad & {x'}_{j,k}^{(\bval)} = x_{j,k}^{(\bval)} + \sum_{i \in [I]}a^{(\bval)}_{i,j,k}y_i \quad (\bval \in \{0, 1\}, ~j \in [J], ~k \in [K^{(\bval)}]),\label{eq:constr1} \\
& {x'}_{j,1}^{(\bval)} = {x'}^{(\bval)}_{\bigl((j + s^{(\bval)}_{k+1} - 1) ~\textrm{mod} ~J \bigr)+1, k+1} \quad (\bval \in \{0, 1\}, ~j \in [J], ~ k \in [K^{(\bval)}-1]),\label{eq:constr2}\\
& x'^{(\bval)}_{j,k} \in \mathbb{Z}_{\ge 0} \quad (\bval \in \{0, 1\}, ~j \in [J], ~ k \in [K^{(\bval)}]),\label{eq:constr3}\\
& y_i \in \mathbb{Z}_{\ge 0} \quad (i \in [I]), \label{eq:constr4}
\end{align}
where $\bm{x}' \coloneqq (x'^{(\bval)}_{j,k})_{\bval \in \bin, (j, k) \in [J] \times [K^{(\bval)}]} \in \mathbb{Z}^{J (K^{(0)}+K^{(1)})}_{\ge 0}$.
The objective function in Eq.~\eqref{eq:obj} represents the total number of inserted $0$s to be minimized.
Eq.~\eqref{eq:constr1} gives the number of 0s for each segment after the $0$-insertion, and Eq.~\eqref{eq:constr2} ensures that sequences with the same output value become cyclically equal after the $0$-insertion.
Eqs.~\eqref{eq:constr3}--\eqref{eq:constr4} restrict the decision variables to non-negative integers. 
The set of optimal solutions is denoted by $\mathcal Z^*(\bm{s, w})$. 

While the above formulation guarantees the security condition, a valid $(*,0)$-SCFO protocol must satisfy the correctness condition as described in Section \ref{ss:SCFO}.
Therefore, for any optimal solution $\bm{z}^{*} \coloneqq (\bm{x'}^{*},~\bm{y}^{*})$ to the optimization problem $\mathrm{P}(\bm{s, w})$, the following condition must hold:
\begin{equation}
{w'_{1}}^{(0)} \not\sim {w'_{1}}^{(1)},\label{eq:confirm}
\end{equation}
where $w_{1}^{(\bval)} \xrightarrow{\bm{u}^*} {w'_{1}}^{(\bval)}$ for $\bval \in \bin$ and $\bm{u}^* = (0^{y^*_1}, 0^{y^*_2}, \ldots, 0^{y^*_{I}})$. 
% Here, $\bm{u}^*$ is 
% ${w'_{1}}^{(\bval)}$ is a 
% \begin{align}
% {x'}_{j,1}^{(0)*} \neq {x'}^{(1)*}_{\bigl((j + s^{(1)}_{1} - 1) ~\mathrm{mod} ~n \bigr)+1, 1} \quad (j \in [n]).\label{eq:confirm}
% \end{align}
This condition means that a sequence with output value 0 and a sequence with output value 1 are not cyclically equal after the $0$-insertion.
Note that since sequences ${w'_{k}}^{(\bval)}$ with the same output value are cyclically equal, it is sufficient to verify this condition for a single pair $({w'_{1}}^{(0)}, {w'_{1}}^{(1)})$.

\subsection{Search Algorithm} \label{ss:algorithm}
In Section~\ref{ss:formulation}, we formulated the search for a $(*,0)$-SCFO protocol as the optimization problem $\mathrm{P}(\bm{s, w})$ given the set of sequences $\bm{w}$ and shift amounts $\bm{s}$.

To verify the existence of a $(*,0)$-SCFO protocol, we present an exhaustive search algorithm that attempts to solve the optimization problem \eqref{eq:obj}–\eqref{eq:constr4} for all permutations of the input sequence bits and all combinations of the shift amounts.
Our algorithm is shown in Algorithm~\ref{alg}.
If our algorithm outputs $\mathcal{F} = \varnothing$, there are no $(*,0)$-SCFO protocols computing $f$.

\begin{algorithm}[t]
\caption{Search Algorithm for $(*,0)$-SCFO Protocols}
\label{alg}
\begin{algorithmic}[1]
\renewcommand{\algorithmicrequire}{\textbf{Input:}}
\renewcommand{\algorithmicensure}{\textbf{Output:}}
\REQUIRE A function $f: \bin^n\ra \bin$
%Initial set of input sequences $\bm{v} = \bm{v}^{(0)} \sqcup \bm{v}^{(1)}$
\STATE Compute the set of input sequences $\bm{v} = \bm{v}^{(0)} \sqcup \bm{v}^{(1)}$ of Eq.~(\ref{eq:def_v}).
%\STATE Calculate from $\bm{w}_0$ the number of 0s $J^{(0)}$, the number of 1s $J^{(1)}$, and the bit length $I$.
%\STATE Generate the set of shift amounts $\mathcal{S}$ using Eq.~\eqref{eq:comb_s}.
\STATE Initialize the set of optimal solutions $\mathcal{F} \leftarrow \varnothing$.
\FOR{$\pi \in \mathfrak{S}_{I}$}
\STATE $\bm{w} \leftarrow \pi(\bm{v})$
\FOR{$\bm{s} \in \{0,1,\ldots,J-1\}^{K^{(0)}+K^{(1)}-2}$}
\IF{the optimization problem $\mathrm{P}(\bm{s},\bm{w})$ is feasible}
\FOR{$\bm{z}^* \in \mathcal{Z}^*(\bm{s},\bm{w})$}
\IF{$\bm{z}^*$ satisfies Eq.~\eqref{eq:confirm}}
\STATE $\mathcal{F} \leftarrow \mathcal{F} \cup \{(\bm{s},\bm{w},\bm{z}^*)\}$
\ENDIF
\ENDFOR
\ENDIF
\ENDFOR
\ENDFOR
\ENSURE The set of optimal solutions $\mathcal{F}$
\end{algorithmic}
\end{algorithm}

\section{Experimental Results}
In this section, we apply Algorithm~\ref{alg} to the 10 representative 3-variable Boolean functions (Nos. 5--14 in the list in Section~\ref{ss:three_variable_function}) and demonstrate the impossibility of $(*,0)$-SCFO protocols for these functions.
The existence of a protocol can be confirmed by finding at least one solution.

To solve the optimization problem defined by Eqs.~\eqref{eq:obj}--\eqref{eq:constr4}, we used the HiGHS 1.8.0 solver, which is integrated into the SciPy library in Python.
All experiments were conducted on a MacBook Air equipped with an Apple M3 processor (8 threads, 4.05 GHz) and 16 GB of RAM.

The exhaustive search found no feasible solution for any of the representative functions, except for the 3-variable equality function (No. 7 in the list in Section~\ref{ss:three_variable_function}).
This result establishes that it is impossible to construct a $(*,0)$-SCFO protocol for the other nine functions.
The computation time for each function ranged from approximately 3 to 60 minutes.

\section{Conclusion}

In this paper, we proposed a new method for proving impossibility results in card-based cryptography based on mathematical optimization. Specifically, we formulated the constructability of $(\ast,0)$-SCFO and $(0,\ast)$-SCFO protocols as an integer optimization problem and proposed a search algorithm. As a result of applying the proposed method to all 3-variable Boolean functions, we demonstrated through computer experiments that no $(\ast,0)$-SCFO protocol exists, except for known protocols (e.g., the equality protocol).

This research has shown the usefulness of mathematical optimization in card-based cryptography. On the other hand, research on card-based cryptography based on mathematical optimization is still in its early stages, and many open problems remain. For example, the following problems are open:
\begin{itemize}
  \item The constructability of $(\ast,0)$-SCFO protocols for Boolean functions with four or more variables is an open problem. Since the computational complexity of the proposed method is at least exponential with respect to the number of variables $n$, applying it directly to 4-variable functions is difficult from a computational complexity standpoint.
  \item While the constructability in this study is for $(\ast,0)$-SCFO and $(0,\ast)$-SCFO protocols, the impossibility of general SCFO protocols that use any number of both $\blk$ and $\red$ as additional cards remains an open problem.
  \item Modeling protocols other than SCFO protocols as integer optimization problems to investigate their constructability is an open problem.
\end{itemize}

\bibliographystyle{abbrv}
\bibliography{card}

\begin{thebibliography}{10}

\bibitem{BoerEC1989}
B.~D. Boer.
\newblock More efficient match-making and satisfiability the five card trick.
\newblock In J.-J. Quisquater and J.~Vandewalle, editors, {\em Advances in
  Cryptology -- EUROCRYPT' 89}, volume 434 of {\em LNCS}, pages 208--217,
  Heidelberg, 1990. Springer.

\bibitem{Cheung2013}
E.~Cheung, C.~Hawthorne, and P.~Lee.
\newblock Cs 758 project: Secure computation with playing cards, 2013.

\bibitem{KilianC1994}
C.~Cr{\'e}peau and J.~Kilian.
\newblock Discreet solitary games.
\newblock In D.~R. Stinson, editor, {\em Advances in Cryptology---CRYPTO' 93},
  volume 773 of {\em LNCS}, pages 319--330, Berlin, Heidelberg, 1994. Springer.

\bibitem{FujitaAPKC2025}
K.~Fujita, S.~Ikeda, K.~Shinagawa, and K.~Yoneyama.
\newblock Formal verification and proof of impossibility for four-card xor
  protocols using only random cuts.
\newblock In K.~Emura and H.~Morita, editors, {\em ACM ASIA Public-Key
  Cryptography Workshop}, New York, 2025. ACM.

\bibitem{HashimotoIEICE2018a}
Y.~Hashimoto, K.~Nuida, K.~Shinagawa, M.~Inamura, and G.~Hanaoka.
\newblock Toward finite-runtime card-based protocol for generating a hidden
  random permutation without fixed points.
\newblock {\em IEICE Trans. Fundam.}, E101.A(9):1503--1511, 2018.

\bibitem{HeatherFAOC2014}
J.~Heather, S.~Schneider, and V.~Teague.
\newblock Cryptographic protocols with everyday objects.
\newblock {\em Formal Aspects Comput.}, 26(1):37--62, 2014.

\bibitem{KastnerAC2017}
J.~Kastner, A.~Koch, S.~Walzer, D.~Miyahara, Y.~Hayashi, T.~Mizuki, and
  H.~Sone.
\newblock The minimum number of cards in practical card-based protocols.
\newblock In T.~Takagi and T.~Peyrin, editors, {\em Advances in
  Cryptology---ASIACRYPT 2017}, volume 10626 of {\em LNCS}, pages 126--155,
  Cham, 2017. Springer.

\bibitem{KochMC2022}
A.~Koch.
\newblock The landscape of optimal card-based protocols.
\newblock {\em Mathematical Cryptology}, 1(2):115--131, 2022.

\bibitem{KochAC2019}
A.~Koch, M.~Schrempp, and M.~Kirsten.
\newblock Card-based cryptography meets formal verification.
\newblock In S.~D. Galbraith and S.~Moriai, editors, {\em Advances in
  Cryptology--ASIACRYPT 2019}, volume 11921 of {\em LNCS}, pages 488--517,
  Cham, 2019. Springer.

\bibitem{KochNGCO2021}
A.~Koch, M.~Schrempp, and M.~Kirsten.
\newblock Card-based cryptography meets formal verification.
\newblock {\em New Gener. Comput.}, 39(1):115--158, 2021.

\bibitem{KochAC2015}
A.~Koch, S.~Walzer, and K.~H{\"a}rtel.
\newblock Card-based cryptographic protocols using a minimal number of cards.
\newblock In T.~Iwata and J.~H. Cheon, editors, {\em Advances in
  Cryptology---ASIACRYPT 2015}, volume 9452 of {\em LNCS}, pages 783--807,
  Berlin, Heidelberg, 2015. Springer.

\bibitem{MarcedoneEPRINT2015}
A.~Marcedone, Z.~Wen, and E.~Shi.
\newblock Secure dating with four or fewer cards.
\newblock Cryptology ePrint Archive, Report 2015/1031, 2015.

\bibitem{MizukiISEC2016}
T.~Mizuki.
\newblock Applications of card-based cryptography to education.
\newblock In {\em Information Security Workshop (ISEC)}, 2016 (in Japanese).

\bibitem{MizukiIJIS2014}
T.~Mizuki and H.~Shizuya.
\newblock A formalization of card-based cryptographic protocols via abstract
  machine.
\newblock {\em Int. J. Inf. Secur.}, 13(1):15--23, 2014.

\bibitem{NiemiTCS1998}
V.~Niemi and A.~Renvall.
\newblock Secure multiparty computations without computers.
\newblock {\em Theor. Comput. Sci.}, 191(1--2):173--183, 1998.

\bibitem{ShinagawaSCIS2022}
K.~Shinagawa.
\newblock A report on a lecture for elementary and junior high school using
  card-based cryptography.
\newblock In {\em 2022 Symposium on Cryptography and Information Security (SCIS
  2022)}, 2022 (in Japanese).

\bibitem{ShinagawaICTAC2024}
K.~Shinagawa.
\newblock Card-based protocols with single-card encoding.
\newblock In C.~Anutariya and M.~M. Bonsangue, editors, {\em Theoretical
  Aspects of Computing}, volume 15373 of {\em LNCS}, pages 182--194, Cham,
  2025. Springer.

\bibitem{ShinagawaICISC2018}
K.~Shinagawa and T.~Mizuki.
\newblock The six-card trick: Secure computation of three-input equality.
\newblock In K.~Lee, editor, {\em Information Security and Cryptology}, volume
  11396 of {\em LNCS}, pages 123--131, Cham, 2018. Springer.

\bibitem{ShinagawaSTACS2025}
K.~Shinagawa and K.~Nuida.
\newblock Card-based protocols imply {PSM} protocols.
\newblock In O.~Beyersdorff, M.~Pilipczuk, E.~Pimentel, and N.~K. Th\`{a}ng,
  editors, {\em Theoretical Aspects of Computer Science}, volume 327 of {\em
  LIPIcs}, pages 72:1--72:18, Dagstuhl, 2025. Schloss Dagstuhl.

\bibitem{ShinagawaFCT2025}
K.~Shinagawa and K.~Nuida.
\newblock Cyclic equalizability of words and its application to card-based
  cryptography.
\newblock In {\em Fundamentals of Computation Theory}, LNCS, Cham, 2025.
  Springer.

\bibitem{ShinagawaARXIV2025}
K.~Shinagawa and K.~Nuida.
\newblock A note on single-cut full-open protocols.
\newblock {arXiv: 2507.03323}, 2025.

\end{thebibliography}

\appendix

\section{Protocol for $x_1\oplus x_2$}

This protocol is a $(0,0)$-SCFO protocol for the $2$-variable XOR function $x_1\oplus x_2$. 
It was proposed by Shinagawa--Mizuki~\cite{ShinagawaICISC2018}.

\begin{enumerate}
    \item The input sequence is given as follows:
    \[
    \underset{x_1}{\back} \, \underset{\ol{x_1}}{\back} \, \underset{x_2}{\back}\, \underset{\ol{x_2}}{\back} \,.
    \]
    \item Apply a random cut to the sequence.
    \item Open all cards. Output $0$ if it is a cyclic shift of $\red\,\blk\,\red\,\blk\,$, and $1$ if it is a cyclic shift of $\red\,\red\,\blk\,\blk\,$.
\end{enumerate}

\section{Protocol for $x_1 \wedge x_2$}

This protocol is a $(0,1)$-SCFO protocol for the $2$-variable AND function $x_1 \wedge x_2$. 
It was proposed by den Boer~\cite{BoerEC1989}.

\begin{enumerate}
    \item The input sequence is given as follows:
    \[
    \underset{\ol{x_1}}{\back} \, \underset{x_1}{\back} \, \underset{\heartsuit}{\back} \, \underset{x_2}{\back}\, \underset{\ol{x_2}}{\back} \,.
    \]
    \item Apply a random cut to the sequence.
    \item Open all cards. Output $0$ if it is a cyclic shift of $\red\,\blk\,\red\,\blk\,\red\,$, and $1$ if it is a cyclic shift of $\red\,\red\,\red\,\blk\,\blk\,$.
\end{enumerate}

\section{Protocol for $(x_1=x_2=x_3)?$}

This protocol is a $(0,0)$-SCFO protocol for the $3$-variable equality function $(x_1=x_2=x_3)?$ that outputs $1$ if and only if $x_1 = x_2 = x_3$. 
It was proposed by Heather--Schneider--Teague~\cite{HeatherFAOC2014} and independently rediscovered by Shinagawa--Mizuki~\cite{ShinagawaICISC2018}. 

\begin{enumerate}
    \item The input sequence is given as follows:
    \[
    \underset{x_1}{\back} \, \underset{\ol{x_2}}{\back} \, \underset{x_3}{\back}\, \underset{\ol{x_1}}{\back} \, \underset{x_2}{\back} \, \underset{\ol{x_3}}{\back} \,.
    \]
    \item Apply a random cut to the sequence.
    \item Open all cards. Output $0$ if it is a cyclic shift of $\red\,\blk\,\red\,\blk\,\red\,\blk\,$, and $1$ if it is a cyclic shift of $\red\,\red\,\red\,\blk\,\blk\,\blk\,$. 
\end{enumerate}

\section{Protocol for $(x_1\wedge x_2) \vee (\ol{x_1} \wedge x_3)$}

This protocol is a $(1,1)$-SCFO protocol for a $3$-variable function $(x_1\wedge x_2) \vee (\ol{x_1} \wedge x_3)$. 
It was proposed by Shinagawa--Nuida~\cite{ShinagawaARXIV2025}. 

\begin{enumerate}
    \item The input sequence is given as follows:
    \[
    \underset{x_2}{\back} \, \underset{\ol{x_1}}{\back}\, \underset{\ol{x_3}}{\back} \, \underset{1}{\back} \, \underset{\ol{x_2}}{\back} \, \underset{x_1}{\back}\, \underset{x_3}{\back} \, \underset{0}{\back} \,.
    \]
    \item Apply a random cut to the sequence.
    \item Open all cards. Output $0$ if it is a cyclic shift of $\red\,\red\,\red\,\red\,\blk\,\blk\,\blk\,\blk\,$, and $1$ if it is a cyclic shift of $\blk\,\blk\,\red\,\blk\,\red\,\red\,\blk\,\red\,$. 
\end{enumerate}

\section{Protocol for $x_1 \oplus x_2 \oplus x_3$}

This protocol is an SCFO protocol for the $3$-variable XOR function $x_1 \oplus x_2 \oplus x_3$. 
It is not standard since the input sequence contains two pairs of $(x_1, \ol{x_1})$. 
It was proposed by Shinagawa--Nuida~\cite{ShinagawaARXIV2025}. 

\begin{enumerate}
    \item The input sequence is given as follows:
    \[
    \underset{x_1}{\back} \, \underset{x_2}{\back}\, \underset{\ol{x_1}}{\back} \, \underset{x_3}{\back} \, \underset{x_1}{\back} \, \underset{\ol{x_2}}{\back}\, \underset{\ol{x_1}}{\back} \, \underset{\ol{x_3}}{\back} \,.
    \]
    \item Apply a random cut to the sequence.
    \item Open all cards. Output $0$ if it is a cyclic shift of $\blk\,\blk\,\red\,\blk\,\blk\,\red\,\red\,\red\,$, and $1$ if it is a cyclic shift of $\red\,\red\,\blk\,\red\,\red\,\blk\,\blk\,\blk\,$. 
\end{enumerate}

\section{Protocol for $\ol{x_1}x_2\ol{x_4} \vee \ol{x_2}x_3\ol{x_4} \vee x_1\ol{x_2}x_4 \vee x_2\ol{x_3}x_4$}

This protocol is a $(0,0)$-SCFO protocol for a $4$-variable function $\ol{x_1}x_2\ol{x_4} \vee \ol{x_2}x_3\ol{x_4} \vee x_1\ol{x_2}x_4 \vee x_2\ol{x_3}x_4$. 
This protocol was proposed by Shinagawa--Nuida~\cite{ShinagawaARXIV2025}. 

\begin{enumerate}
    \item The input sequence is given as follows:
    \[
    \underset{x_1}{\back} \, \underset{x_2}{\back}\, \underset{x_3}{\back} \, \underset{x_4}{\back} \, \underset{\ol{x_1}}{\back} \, \underset{\ol{x_2}}{\back}\, \underset{\ol{x_3}}{\back} \, \underset{\ol{x_4}}{\back} \,.
    \]
    \item Apply a random cut to the sequence.
    \item Open all cards. Output $0$ if it is a cyclic shift of $\red\,\red\,\red\,\red\,\blk\,\blk\,\blk\,\blk\,$, and $1$ if it is a cyclic shift of $\blk\,\blk\,\red\,\blk\,\red\,\red\,\blk\,\red\,$. 
\end{enumerate}

\end{document}